\documentclass{ws-p9-75x6-50}

\begin{document}

\title{Relativity of motion in quantum vacuum}

\author{Marc-Thierry Jaekel}

\address{Laboratoire de Physique Th\'eorique
\\Ecole Normale Sup\'{e}rieure\\
24, rue Lhomond, F-75231 Paris Cedex 05, France}

\author{Astrid Lambrecht and Serge Reynaud}

\address{Laboratoire Kastler Brossel
\footnote{mailto:reynaud@spectro.jussieu.fr; 
http://www.spectro.jussieu.fr/Vacuum}\\
Universit\'{e} Pierre et Marie Curie, Ecole Normale Sup\'{e}rieure et CNRS\\
Campus Jussieu, Case 74, F-75252 Paris Cedex 05, France}

\maketitle

\abstracts{The problem of relativity of motion in quantum vacuum is addressed
by considering a cavity moving in vacuum in a monodimensional space. 
The cavity is an open system which emits photons when it oscillates in vacuum. 
Qualitatively new effects like pulse shaping in the time domain and frequency 
conversion in the spectrum may help to discrimate motion-induced radiation
from potential stray effects.}

\section{Introduction}

Relativity of motion is one of the basic principles of physics since Galileo. In 
classical physics this principle applies to motion in vacuum which is just another 
name for empty space. But the face of the problem today is changed by quantum 
theory. Quantum vacuum is no longer empty. It contains field fluctuations 
which lead to mechanical effects for any scatterer in vacuum. In this paper we will 
discuss some observable effects associated with the motion of mirrors in vacuum.

Here we restrict our attention to the vacuum fluctuations of the electromagnetic 
field. These fluctuations are known to exert a mechanical action on scatterers. 
Their coupling to electrons in atoms lead to phenomena like spontaneous emission and 
the Lamb shift of energy levels for a single atom, or van der Waals forces between 
two atoms or molecules. 

For macroscopic objects, the most famous effect induced by vacuum fluctuations is 
the Casimir force arising between two mirrors in vacuum \cite{Casimir48}. 
For a single mirror moving in vacuum there also exists a dissipative force opposing 
itself to the mirror's motion \cite{FullingDavies76}. Even when the mirror is at rest 
in vacuum, it experiences a fluctuating force due to the radiation pressure of field 
fluctuations \cite{Barton91,Jaekel92a}. The dependence of the dissipative force is
directly connected to the spectral properties of the fluctuating force
through the fluctuations-dissipation relations \cite{Jaekel97}. 

\section{The dissipative force}

Let us begin with the simple model of a perfect mirror in a two-dimensional 
space-time. In a thermal field, the dissipative force $F_{\rm diss}(t)$
is proportional to the mirror's velocity $q^\prime(t)$ 
\begin{equation}
F_{\rm diss}(t) = -\frac{\hbar \theta^2}{6 \pi c^2} q^\prime(t)
\end{equation}
The force may equivalently be written in the frequency domain 
\begin{equation}
F_{\rm diss}[\omega] = \frac{\hbar \theta^2}{6 \pi c^2} i\omega q[\omega]
\label{Fdissclass}
\end{equation}
where $F_{\rm diss}[\omega]$ and $q[\omega]$ are the Fourier transform of the 
force and mirror's displacement. In both formulas, $\theta$ is the field temperature 
expressed in frequency units
\begin{equation}
\theta = \frac{2 \pi k_{\rm B} T_{\rm field}}{\hbar}
\end{equation}
$\hbar$, $k_{\rm B}$ and $c$ are the Planck constant, the Boltzmann constant 
and the speed of light respectively. This force is a classical expression which 
tends towards zero when temperature goes to zero. In fact, it neglects the effect
of vacuum fluctuations.

When this effect is taken into account, the linear susceptibility is found to 
scale as the third power of frequency at the limit of zero temperature
\begin{equation}
F_{\rm diss}[\omega] = \frac{\hbar}{6 \pi c^2}i\omega^3 q[\omega]
\label{Fdiss}
\end{equation}
This result, which could be expected from mere dimensional arguments,
implies that the force is proportional to the third order time derivative 
of the mirror's position 
\begin{equation}
F_{\rm diss}(t) = \frac{\hbar}{6 \pi c^2} q^{\prime\prime\prime}(t)
\end{equation}
The linear susceptibilities (\ref{Fdissclass},\ref{Fdiss}) are directly connected 
to the spectral properties of the fluctuating force exerted upon a mirror
at rest through the fluctuations-dissipation relations \cite{Jaekel97}. 
At an arbitrary temperature the dissipative force is just the sum of the 2
contributions (\ref{Fdissclass}) and (\ref{Fdiss}). These expressions can also be 
generalized to the case of a real mirror with frequency-dependent reflection and 
transmission amplitudes \cite{Jaekel93}. 

The dissipative force arising for a mirror moving in quantum vacuum has interesting 
consequences with respect to the problem of relativity of motion. This force 
vanishes for a uniform velocity, as expected from the Lorentz invariance of quantum 
vacuum. It also vanishes for a uniform acceleration. The appearance of vacuum in an 
accelerated 
frame is a much debated question \cite{Jaekel98}. For the present problem of 
motion of a mirror in vacuum we have a clear answer at our disposal. 
No dissipative force arises for a motion with uniform acceleration and this fact may 
be explained as a consequence of the conformal invariance of electromagnetic vacuum. 

Now, there exists a non-vanishing dissipative force for a mirror moving in vacuum 
with a non-uniform acceleration. This means that vacuum fluctuations have observable 
mechanical effects on a mirror moving without any further reference than vacuum 
fluctuations themselves. In other words, this implies that quantum vacuum may be 
considered as defining privileged reference frames for motion. 

\section{Observation of dissipative effects?}

The dissipative effects related to motion in vacuum have never been observed for 
macroscopic objects like mirrors. As a matter of fact, the orders of magnitude
are exceedingly small for the fluctuating force as well as for the dissipative 
force. This raises the question which we discuss in this paper: how can one 
observe dissipative effects of vacuum fluctuations on mirrors? \cite{Lambrecht96}

\begin{figure}[htb]
\centerline{\epsfig{figure=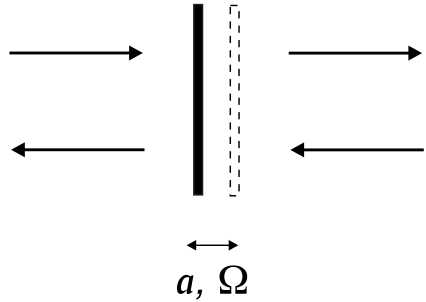,height=4cm}}
\caption{Single mirror oscillating in vacuum. The arrows represent the vacuum 
field which, 
in a monodimensional space, may be considered as two counterpropagating fields.}
\label{fig1}
\end{figure}

A first idea is to observe changes in the field rather than in the mechanical 
forces. Indeed, due to energy conservation photons are emitted into vacuum 
when the mirror's motion is damped. In other words, the dissipated energy is 
transformed into radiation. Let us consider a mirror oscillating in vacuum at 
a frequency $\Omega$ and with an amplitude $a$ (cf. figure \ref{fig1}).
The number of emitted photons $N$ during the measurement time $T$ can then be 
written in the following way 
\begin{eqnarray}
N &=& \frac{\Omega^3 a^2 T}{3 \pi c^2} = \frac{\Omega T}{3\pi}\frac{v^2}{c^2} 
\nonumber \\
v &=& a \Omega   
\label{N1} 
\end{eqnarray}
$\Omega^3$ characterizes the already discussed motional susceptibility and 
$v$ is the mirror's maximal velocity. Since $N$ scales as the square of 
the ratio between the mirrors mechanical velocity and the speed of light,
it remains very small for any possible macroscopic motion. If we 
consider a macroscopic velocity to be bound by the sound velocity in typical 
materials (e.g. quartz), one obtains at most one emitted photon per $10^{10}$ 
oscillation periods. 

A second idea for improving the orders of magnitude of this motion-induced radiation 
is to study a cavity oscillating in vacuum instead of a single mirror 
\cite{Lambrecht96}. This configuration allows one to profit from the resonant 
amplification of the opto-mechanical coupling between the field and the moving 
mirrors \cite{Jaekel92}.
The resonant enhancement is determined by the cavity finesse ${\cal F}$ which gives 
the number of roundtrips of the field before it leaves the cavity.
Hence the cavity has to be treated as an open system with mirrors having reflection 
coefficients smaller than unity so that the field can escape it by transmission 
through the mirrors. This distinguishes these calculations from the numerous works 
devoted to photon production between a pair of perfectly reflecting mirrors 
\cite{FullingDavies76,PerfectMirrors} in which case the amount of radiation 
emitted outside the cavity cannot be evaluated. 

We have shown \cite{Lambrecht96} that the motion-induced radiation is effectively 
enhanced under opto-mechanical resonance conditions.
\begin{figure}[htb]
\centerline{\epsfig{figure=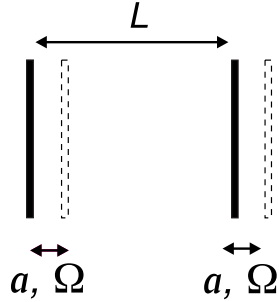,height=4cm}}
\caption{Cavity oscillating globally in vacuum with an amplitude $a$ at a frequency 
$\Omega$.}
\label{fig2}
\end{figure}
For a motion of the cavity as a whole (see figure \ref{fig2}), 
this occurs when the mechanical frequency $\Omega$ is an odd multiple of the 
fundamental cavity resonance frequency $\pi/\tau$
\begin{eqnarray}
\Omega &=& \frac{3\pi}{\tau}, \frac{5\pi}{\tau}, \frac{7\pi}{\tau}, \ldots 
\nonumber \\
\tau &=& \frac L c
\end{eqnarray}
$\tau$ is the time of flight of photons between the two mirrors separated by a 
distance $L$. At perfectly tuned resonance, the number of photons emitted by the 
cavity is the product of the number (\ref{N1}) corresponding to a single mirror by 
the cavity finesse ${\cal F}$ 
\begin{eqnarray}
N &=& {\cal F}\ \frac{\Omega T}{3\pi} \frac{v^2}{c^2}
\end{eqnarray}
Since the cavity finesse can be a very large number, up to $10^9$ for instance for 
microwave cavities, this makes the experimental observation of motion-induced 
radiation a much less difficult challenge.

In order to make this experimental observation feasible, we have also looked for
signatures which might help to discriminate motion induced radiation from potential 
stray effects. To this aim, we have performed a detailed study of the temporal and
spectral features of this radiation \cite{Lambrecht98}.

\section{Pulse shaping}

\begin{figure}[htb]
\centerline{\epsfig{figure=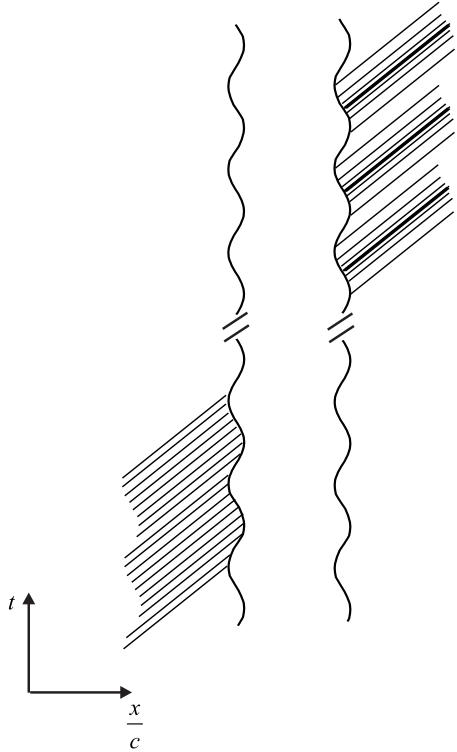,height=10cm}}
\caption{Space-time diagram of the multiple scattering process for a cavity 
oscillating as a whole in vacuum}
\label{fig3}
\end{figure}

The multiple scattering of the field by the moving cavity is represented on the 
space-time diagram of figure \ref{fig3}. 
On this diagram light rays are presented by lines making a 45 degree angle with the 
space and the time axis. The moving mirrors correspond to the sinusoidal lines. 
The scale of the mirrors oscillations is largely exaggerated. 

\begin{figure}[htb]
\centerline{\epsfig{figure=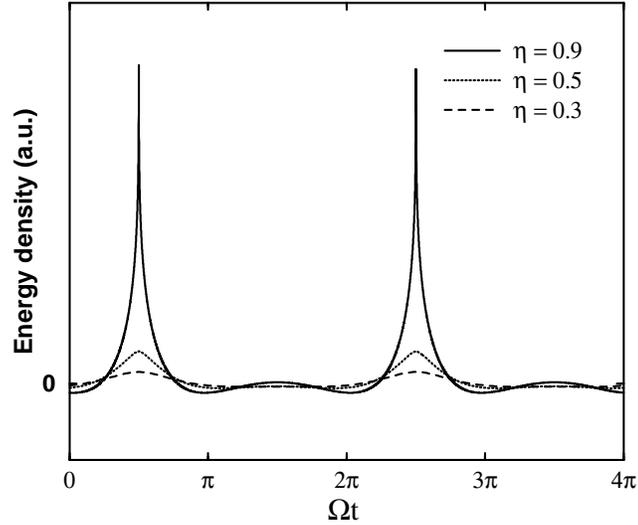,height=8cm}}
\caption{Energy density emitted by the cavity as a function of time}
\label{fig4}
\end{figure}

The point is that multiple scattering gives rise to periodic orbits. An incoming 
light ray is attracted to the neighboring stable orbit while it is repelled from 
the neighboring unstable orbit. When considering a fixed number of scattering 
processes one obtains the input-output transformation shown in figure \ref{fig3}.
This process leads to the formation of regularly spaced field pulses bouncing back 
and forth the cavity. At each scattering on one of the mirrors, there is a 
small probability for a photon for escaping the cavity and therefore being detected 
outside the cavity. This probability is given by the inverse of the cavity finesse. 
Figure \ref{fig4} shows the energy density into vacuum emitted by the oscillating 
cavity as a function of time. 
We have plotted the energy density for three different values of $\eta$,
which is the ratio of the effective phase velocity ${\cal F} v$ 
characterizing the efficiency of the multiple scattering to the velocity of light
\begin{equation}
\eta = {\cal F} \frac{v}{c}
\end{equation}
As already discussed, the single scattering parameter $\frac{v}{c}$ is necessarily
very small for macroscopic motions but this is not the case for the multiple 
scattering parameter $\eta$ thanks to the multiplication by the cavity finesse. 
Figure \ref{fig4} shows that the pulses become sharp and high when the effective 
phase velocity approaches the velocity of light. The plot is based on an analytical 
solution of the multiple scattering problem in terms of homographic mappings of 
phase exponentials \cite{Lambrecht98}. This approach remains valid in the case of 
interest $\eta \sim 1$ whereas a linearized approach would be restricted to 
$\eta \ll 1$.

\section{Frequency conversion}

\begin{figure}[htb]
\centerline{\epsfig{figure=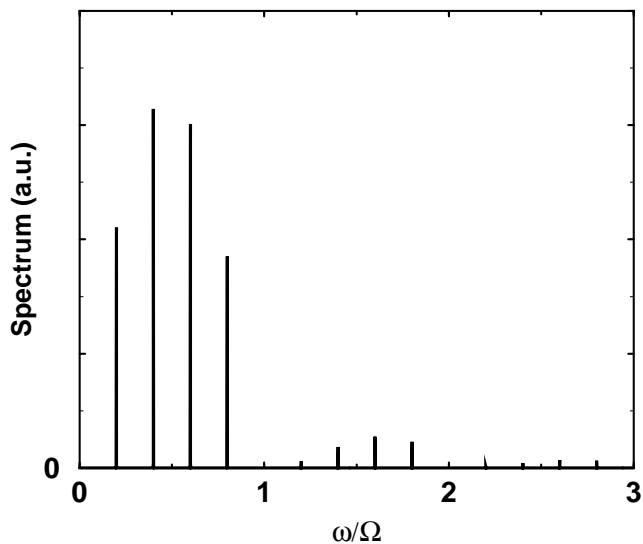,height=8cm}}
\caption{Photon number spectrum of the emitted radiation for $\eta=0.9$ as a 
function of frequency $\omega$ normalized by the mechanical oscillation frequency 
$\Omega$.}
\label{fig5}
\end{figure}

Another interesting feature is the frequency spectrum of the emitted 
radiation shown in figure \ref{fig5}. 
Radiation is emitted at the resonance frequencies of the cavity. The spectrum shown 
here is plotted for a cavity oscillating at a frequency 
\begin{equation}
\Omega = \frac{5\pi}{\tau}
\end{equation}
This means that the cavity performs five oscillations during one roundtrip of
the field inside the cavity. Photons are emitted at multiple integers of the 
fundamental cavity frequency, that is at specific rational multiples of the 
mechanical excitation frequency $\Omega$
\begin{eqnarray}
\omega &=& \frac{\pi}{\tau}, \frac{2\pi}{\tau}, \frac{3\pi}{\tau}, 
\frac{4\pi}{\tau}, \frac{6\pi}{\tau}, \ldots \nonumber \\
&=& \frac{\Omega}{5}, \frac{2\Omega}{5}, \frac{3\Omega}{5}, \frac{4\Omega}{5},
\frac{6\Omega}{5}, \ldots
\end{eqnarray}
A striking feature is that no radiation is emitted at multiple integers of $\Omega$. 
In addition, photons are emitted not only for frequencies lower but also for 
frequencies higher than the oscillation frequency $\Omega$. 

\section{Orders of magnitude}

Clearly, the specific temporal and spectral signatures of the emission may help to 
discriminate motion-induced radiation from potential stray effects in an 
experimental observation. 

To be more specific about the orders of magnitude, let us recall that we
have assumed the input fields to be in the vacuum state. This assumption
requires the number of thermal photons per mode to be smaller than $1$ in
the frequency range of interest
\begin{equation}
\hbar \omega \ll k_{\rm B} T 
\end{equation}
Low temperature requirements thus point to experiments using small mechanical 
structures with optical resonance frequencies as well as mechanical oscillation
frequencies in the {\rm GHz} range. This corresponds to an operation temperature 
\begin{equation}
T \sim 10 {\rm mK}
\end{equation}
In these conditions, the finesse of a superconducting cavity \cite{Rydberg} 
can reach $10^9$. A peak velocity 
\begin{equation}
v \sim 0.3 {\rm m/s}
\end{equation}
would thus be sufficient to obtain a multiple scattering parameter $\eta$
close to unity. The radiated flux of $10$ photons per second outside the cavity 
could be detected by efficient photon-counting detection available
in the {\rm GHz} range. Alternatively, the photons produced inside the cavity 
could be probed with the help of Rydberg atoms \cite{Rydberg}.

It is important to emphasize that the peak velocity considered here
is only a small fraction of the typical sound velocity in materials
so that fundamental breaking limits do not oppose to these numbers.
This velocity corresponds to a small amplitude 
\begin{equation}
\frac v\Omega \sim 10^{-11} {\rm m}
\end{equation}
but to a very large acceleration
\begin{equation}
\Omega v \sim 10^{10} {\rm m/s}^2
\end{equation}

The observation of motional radiation in vacuum seems to be achievable
by an experiment of this kind. The difficulty remains to find means
for exerting a very large force to excite the motion 
of the cavity while keeping the optical part of the experiment 
at a very low temperature and unaffected by the
stray fields induced by the excitation.

\end{document}